\documentclass[journal, twoside]{IEEEtran}
\usepackage{cite}

\ifCLASSINFOpdf
\else
\fi
\usepackage{amsmath}
\usepackage{multirow}

\usepackage[hyphens]{url}
\usepackage{hyperref}
\usepackage[hyphenbreaks]{breakurl}

\begin{document}
%
\title{Uncovering the Viral Nature of Toxicity in Competitive Online Video Games}
%
%
%

\author{Jacob Morrier, Amine Mahmassani, and R. Michael Alvarez
\thanks{J. Morrier and R. M. Alvarez are with the Division of the Humanities and Social Sciences, California Institute of Technology, Mail Code 228-77, Pasadena, CA, 91125, USA (e-mail: jmorrier@caltech.edu; rma@hss.caltech.edu).}
\thanks{A. Mahmassani is with Activision\textregistered.}%
}

\maketitle

\begin{abstract}
    Toxicity is a widespread phenomenon in competitive online video games. In addition to its direct undesirable effects, there is a concern that toxicity can spread to others, amplifying the harm caused by a single player's misbehavior. In this study, we estimate whether and to what extent a player's toxic speech spreads, causing their teammates to behave similarly. To this end, we analyze proprietary data from the free-to-play first-person action game \emph{Call of Duty\textregistered: Warzone\texttrademark}. We formulate and implement an instrumental variable identification strategy that leverages the network of interactions among players across matches. Our analysis reveals that all else equal, all of a player's teammates engaging in toxic speech increases their probability of engaging in similar behavior by 26.1 to 30.3 times the average player's likelihood of engaging in toxic speech. These findings confirm the viral nature of toxicity, especially toxic speech, in competitive online video games.
\end{abstract}


%
\IEEEpeerreviewmaketitle

\section{Introduction}

\IEEEPARstart{C}{ompetitive} online video games are a popular form of entertainment. They allow players to immerse themselves in virtual environments and interact with other players. While these games offer a positive experience to many, they can also expose players to disruptive behavior, including bullying, cheating, trolling, and toxicity.

Previous research suggests toxicity is so deeply ingrained in competitive online video games that it has become normalized, with some players rationalizing it as an inherent part of the game culture \cite{turkay_2020, beres_2021}. For instance, a recent study found that 86\% of players aged 18 to 45 report having experienced harassment while playing \cite{adl_2022}. The incidence of toxic behavior in competitive online video games can be accounted for by their inherently competitive nature, amplified by the anonymity of online interactions \cite{LAPIDOTLEFLER2012434, adinolf_turkay_2018, Lee:2019:0301-2212:1}. Given these games' massive player bases, even a low rate of toxic behavior translates into thousands of daily incidents, exposing an even higher number of players to such behavior.

The undesirable effects of toxicity on player performance, psychological well-being, satisfaction, and retention are well-documented \cite{kwak_blackburn_han_2015}. For example, players who are frequently targeted by toxic behavior tend to report more symptoms of depression and problematic gaming habits \cite{CAPLAN20091312, zsila_2022}. Furthermore, both victims and perpetrators tend to report heightened levels of anxiety and anger rumination. Marginalized groups, such as LGBTQ+ individuals, people of color, and women, are disproportionately affected by toxicity and its detrimental effects \cite{gray_2012, salter_blodgett_2012, kuznekoff_rose_2013, FOX2014314, chess_shaw_2015, ballard_2017, madden_2021}. In esports, toxic behavior has been found to disrupt team coordination and morale, ultimately hindering players' in-game performance \cite{turkay_2020}. Lastly, it has been shown that toxicity contributes to player churn and deters new players from joining, suggesting that video game publishers have a vested interest in mitigating toxicity \cite{kowert_kilmer}.

In addition to these first-order effects, there is a concern that toxic behavior can spread to other players, amplifying the consequences of a single player's misbehavior. Countless empirical studies, experimental and observational, have documented strong correlations and causal relationships between an individual's behavior and outcomes and those of their peers, a phenomenon known as ``peer effects'' \cite{Manski_2000, ALEXANDER200122, SALMIVALLI2010112, EPPLE20111053, Kreager_et_al_2011, SACERDOTE2011249, Graham_2018}. This is equally true of virtuous and reprehensible behavior, including academic dishonesty, bullying, and delinquency. Some studies have specially considered peer effects in online interactions, including how a player's social network influences their performance in multiplayer video games \cite{shen_monge_williams_2014, https://doi.org/10.1111/poms.13007, 10.1145/3490486.3538343}.

Peer effects can be theoretically justified through several mechanisms, one of which is conformity to social norms \cite{HENRICH1998215}. As inherently social beings, humans tend to avoid deviating from social norms. Accordingly, if toxicity is perceived as consistent with these norms, as suggested by its frequency and players' stated perceptions, players may feel compelled or encouraged to engage in such behavior. In this context, the actions of a single individual have the transformative power to redefine the norm, making behavior once considered unacceptable seem acceptable \cite{10.1145/3411764.3445157}. Therefore, it is plausible to expect toxicity to propagate, with one player's toxic behavior causing the exposed players to engage in similar behavior. Previous research has already provided correlational evidence in support of this conjecture \cite{SHEN2020106343, kordyaka_2023}. Relatedly, conformity (or lack thereof) has previously been recognized as a factor in why female and other ``out-group'' players may be targeted by toxic behavior \cite{FOX2014314, kwak_blackburn_han_2015}.

Understanding how toxicity presumedly spreads is crucial for accurately appreciating its impact, as this impact mechanically grows with the extent to which it spreads to other players. This information is also valuable for developing effective strategies to address toxicity. To our knowledge, no study has yet quantified the causal relationship between a player's toxic behavior and that of their teammates. In this context, this paper seeks to assess whether and, if so, to what extent toxic behavior, specifically toxic speech, by a player causes their teammates to engage in similar behavior in competitive online video games.

We analyze proprietary data from the first-person action video game franchise \emph{Call of Duty\textregistered}, published since 2003 by Activision\textregistered. We focus on the free-to-play first-person action video game \emph{Call of Duty: Warzone\texttrademark} \cite{warzone}. In its primary game mode, Battle Royale (BR), players compete to be the last player or team standing by eliminating all their opponents. Players can play alone, in pairs, or in squads of three or four players. Matches begin with players skydiving from an aircraft and landing on a large map within a limited time frame. Players start with minimal equipment and must search the map for weapons and gear scattered throughout, all while avoiding being eliminated by enemies. Players can also loot equipment from eliminated players. As the game progresses and more players are eliminated, the playable area shrinks, forcing survivors into smaller zones.

Since 2023, Activision has partnered with Modulate\texttrademark, a start-up developing advanced voice technology to detect and mitigate online toxicity \cite{toxmod_description}. Activision has integrated Modulate's voice chat moderation technology, ToxMod\texttrademark, into its gaming platforms. ToxMod uses artificial intelligence and machine learning to detect toxic speech, including harassment, hate speech, and discriminatory remarks, in real-time \cite{kowert_woodwell}. This technology was rolled out in North America on August 30, 2023, within the games \emph{Call of Duty: Modern Warfare\textregistered II} \cite{modernwarfare} and \emph{Call of Duty: Warzone\texttrademark}, with initial support in English and plans to extend to additional languages. Our analysis is based on the data generated by ToxMod for all matches monitored between September 1 and October 31, 2023. A player is classified as having engaged in toxic speech in a game if ToxMod detected at least one instance of toxicity during it.

Even with a large amount of high-quality data, analysts seeking to estimate the causal effect of toxic speech by a player on their teammates' probability of employing similar language face considerable statistical challenges. The reason is that some variables not controlled for in our regression models---because they are unmeasured, for instance---may simultaneously influence whether players and their teammates engage in toxic speech. For example, teammates might concomitantly use toxic language in reaction to a random event occurring in a match. More fundamentally, players mutually influence each other. Consequently, whether a player and their teammates engage in toxicity is jointly determined. This obscures the cause-to-effect relationship of exposure to toxic speech and introduces biases in standard ordinary least squares (OLS) estimates.

To address this causal identification problem, we propose a novel identification strategy that takes advantage of the fact that we observe players in multiple matches with different teammates and the resulting network of interactions. By implementing an instrumental variable or two-stage least squares (2SLS) estimation strategy, we isolate variations in players' probability of engaging in toxic speech caused by interactions with teammates who, in other matches with other players, have employed toxic language more frequently and are therefore more likely to do so in the current game. This approach allows us to rigorously determine whether, and to what extent, a player's use of toxic speech \emph{causes} variations in their teammates' probability of doing so. In and of itself, our identification strategy represents a significant contribution to the methodological literature on the causal identification of peer effects.

We perform our analysis on four sets of observations: (i)~all observations from the BR Duos game mode, in which players compete in pairs; (ii)~algorithmically matched pairs in BR Duos games; (iii)~all observations in the BR Quads mode, in which players compete in four-member squads; and (iv)~all observations from the BR Duos, Trios and Quads modes. Our findings confirm that toxicity is viral, with players having a significant causal effect on their teammates' likelihood of using toxic language. When all of a player's teammates engage in toxic speech, their probability of engaging in similar behavior rises by 2.91 to 7.57 percentage points (pp.). Depending on the context, this effect represents 26.1 to 30.3 times the average probability of a player using toxic language. These findings confirm the viral nature of toxicity, especially toxic speech, in competitive online video games.

\section{Methodology}

\begin{table*}
    \centering
    \caption{Descriptive Statistics}
    \label{tab:ds}
    \makebox[\linewidth][c]{
    \begin{tabular}{lcccc}
        \hline
        \hline
        \\[-1.4ex]
        & \multirow{2}{*}{\textbf{Duos Mode}} & \textbf{Duos Mode} & \multirow{2}{*}{\textbf{Quads Mode}} & \multirow{2}{*}{\textbf{All Modes}} \\ 
        & & \textbf{Algorithmic Pairs} & & \\
        \\[-1.4ex]
        \hline
        \\[-1.4ex]
        Number of Unique Matches & 244,182 & 240,367 & 303,360 & 605,664 \\
        Number of Unique Players & 1,038,893 & 339,820 & 1,166,445 & 1,963,084 \\
        Number of Observations & 11,958,686 & 2,264,728 & 9,381,343 & 19,131,444 \\
        Average Probability of Toxic Speech (\%) & 0.264 & 0.102 & 0.249 & 0.249 \\
        \\[-1.4ex]
        \hline
        \hline
    \end{tabular}}
\end{table*}

The following methodological explanation focuses on the BR Duos game mode, in which players compete in pairs. We then discuss extending this methodology to the BR Trios and Quads games modes, in which players participate in three- and four-member squads, respectively.

For reference, Table \ref{tab:ds} contains the number of unique matches, unique players, and observations, along with the average probability that a player engages in toxic speech, for each of the four sets of observations on which we conduct our analysis. Our dataset does not contain personally identifiable information permitting the individual to whom the information applies to be reasonably inferred, nor any data on players' sex, gender, race, ethnicity, or other socially relevant groups. Caltech's Committee for the Protection of Human Subjects reviewed and approved our research design.

\subsection{Model Specification}

We wish to estimate the causal effect of a player employing toxic speech on their teammate's likelihood of exhibiting similar behavior. We define this effect as the change in a player's probability of engaging in toxic speech in the current match induced by their teammate engaging in toxic speech, holding all other factors constant.

To this end, we formulate the following structural model of players' behavior:
\[
    Y_{ij} = \alpha_{j} + \beta \times Y_{ik} + \varepsilon_{ij}.
\]

In this model, $Y_{ij}$ is a binary variable denoting whether player $j$ has engaged in toxic speech in match $i$, $\alpha_{j}$ a player-specific intercept reflecting player $j$'s natural tendency to employ toxic language, and $\varepsilon_{ij}$ an error term. We classify a player as having engaged in toxic speech during a match if ToxMod\texttrademark labeled at least one of their statements as toxic.

This model postulates that whether a player engages in toxic speech is determined by two factors: (i)~their inherent predisposition toward such behavior, and (ii)~whether their teammate engages in it. The coefficient $\beta$ reflects the causal effect of a teammate engaging in toxic speech on a player's likelihood of exhibiting similar behavior. This coefficient is our estimand.

\subsection{Causal Identification Problem} 

Naturally, one might contemplate estimating the coefficient~$\beta$ through OLS. However, contrary to standard assumptions in linear regression models, the model's equations are not exclusively linked through their error terms. The reason is that the dependent variable in some equations appears on the right-hand side of others. This entails that a player and their teammate mutually influence each other, and whether they each employ toxic speech is jointly determined. Consequently, the variable indicating whether one's teammate engages in toxic speech is endogenous, resulting in a causal identification problem.

To demonstrate this formally, we consider the pair formed by players $j$ and $k$ in match $i$. The two equations governing whether these players engage in toxic speech are defined as follows:
\begin{equation*}
    \begin{split}
        Y_{ij} & = \alpha_{j} + \beta \times Y_{ik} + \varepsilon_{ij} \\
        Y_{ik} & = \alpha_{k} + \beta \times Y_{ij} + \varepsilon_{ik}.
    \end{split}
\end{equation*}
    
To show that $Y_{ik}$ is correlated with $\varepsilon_{ij}$, it suffices to incorporate the first equation into the second and isolate $Y_{ik}$ on the left-hand side:
\begin{equation*}
    \begin{split}
        & Y_{ik} = \alpha_{k} + \beta \times \left(\alpha_{j} + \beta \times Y_{ik} + \varepsilon_{ij} \right) + \varepsilon_{ik} \\
        \Leftrightarrow & \left(1 - \beta^{2}\right) \times Y_{ik} = \alpha_{k} + \beta \times \left(\alpha_{j} + \varepsilon_{ij}\right) + \varepsilon_{ik} \\
        \Leftrightarrow & Y_{ik} = \frac{\beta}{1 - \beta^{2}} \times \left(\alpha_{j} + \varepsilon_{ij}\right) + \frac{1}{1 - \beta^{2}} \times \left(\alpha_{k} + \varepsilon_{ik}\right).
    \end{split}
\end{equation*}
    
The last equation implies that the error term $\varepsilon_{ij}$ implicitly enters the value of $Y_{ik}$, such that they are correlated. This means that OLS estimates do not only capture the effect of the teammate's toxic speech on a player's probability of engaging in such behavior but also its ``reflection,'' that is, the effect that player has on their teammate.

Other challenges to the causal identification of peer effects include misspecification and self-selection. We discuss each in turn. First, our model may be incorrectly specified, as it may not contain variables governing players' probability of engaging in toxic speech. Some of these factors may equally influence whether a player and their teammate engage in toxicity. For example, two players in a team might engage in toxic behavior after observing one of their opponents doing so or in reaction to their actions. Formally, this may introduce a correlation between the error terms of two players, say, players~$j$ and~$k$, in the same team within a particular match, say, match~$i$:
\[
    \text{corr}\left(\varepsilon_{ij}, \varepsilon_{ik}\right) \not= 0.
\]

Second, players can endogenously choose their teammates by forming ``parties.'' Players may form parties for various reasons, including to engage intentionally in toxicity or in anticipation that their teammate will engage in such behavior. Also, when two players decide to play as a pair, it suggests a certain degree of familiarity between them. This familiarity can affect social dynamics between them. Formally, this may introduce a correlation between the error terms of two players across all matches in which they are paired:
\begin{multline*}
    \text{corr}\left(\varepsilon_{ij}, \varepsilon_{\ell k}\right) \not= 0 \text{ for all matches } i, \ell \text{ in which} \\
    \text{players } j \text{ and } k \text{ are paired.}
\end{multline*}

\subsection{Identification Strategy}

To address the causal identification problem described above, we formulate an identification strategy that leverages the fact that we observe players in multiple matches with different teammates. We propose to implement an instrumental variable or 2SLS estimation strategy. Concretely, we will use as an instrument of the variable indicating whether a player's teammate engaged in toxic speech the probability that they engage in such behavior in other matches in which they are teamed up with other players.

Formally, our identification strategy consists of adding the following equation to our structural model of players' behavior:
\[
    Y_{ik} = \delta_{j} + \gamma \times \Bar{Y}_{k,-j} + u_{ik},
\]
where $\Bar{Y}_{k,-j}$ equals the probability that player $k$ engages in toxic speech in all matches in which they are not paired with player $j$. This instrumental variable belongs to the general class of spatial or ``leave-one-out'' instruments used in empirical industrial organization for demand and supply estimation \cite{Hausman_1996, nevo_2001}. In short, our identification strategy isolates variations in a player's probability of engaging in toxic speech resulting from being paired with a teammate who, in other matches played with other players, has a greater inclination to engage in such behavior.

\begin{table*}
    \centering
    \caption{First-Stage Estimation Results}
    \label{tab:first_stage}
    \makebox[\linewidth][c]{
    \begin{tabular}{lcccc}
        \hline
        \hline
        \\[-1.4ex]
         & \multirow{2}{*}{\textbf{Duos Mode}} & \textbf{Duos Mode} & \multirow{2}{*}{\textbf{Quads Mode}} & \multirow{2}{*}{\textbf{All Modes}} \\ 
         & & \textbf{Algorithmic Pairs} & & \\
         \\[-1.4ex]
         \hline 
         \\[-1.4ex]
        Estimate & 0.2200 & 0.2044 & 0.2575 & 0.1693 \\
         Standard Error & (0.005) & (0.009) & (0.003) & (0.003) \\
         \\[-1.4ex]
         \hline
         \\[-1.4ex]
         $F$ Statistic & 1937.6 & 497.6 & 6855.3 & 1485.4 \\
         Number of Observations & 11,958,686 & 2,264,728 & 9,381,343 & 19,131,444 \\
         \\[-1.4ex]
         \hline
         \hline
    \end{tabular}}
\end{table*}

In general, for an instrumental variable to be valid, it must satisfy two conditions: (i) relevance, meaning that the instrumental variables must be strongly correlated with the endogenous explanatory variables, and (ii) exclusion, meaning that the instrumental variables must be independent of the structural equation’s error term. The validity of the first condition can be verified empirically by examining the first-stage regression. As a rule of thumb, the $F$ statistic against the null hypothesis that the instrument is irrelevant in the first-stage regressions should have a value greater than ten. Table \ref{tab:first_stage} presents the coefficient associated with the instrumental variable, the $F$ statistic, and the number of observations for all first-stage regressions in our analysis. In each case, the $F$ statistic is substantially greater than ten, suggesting that we undoubtedly have a ``strong first stage.''

On the other hand, the validity of the exclusion restriction cannot be empirically tested. Instead, it hinges on the assumptions we are willing to make regarding the relationship between the instrumental variables and the structural equation's error term. To show this, we consider the equation governing whether player $j$ engages in toxic behavior in match $i$, in which they are paired with player $k$:
\[
	Y_{ij} = \alpha_{j} + \beta \times Y_{ik} + \varepsilon_{ij}.
\]

The corresponding first-stage equation is defined as follows:
\[
	Y_{ik} = \delta_{j} + \gamma \times \Bar{Y}_{k,-j} + u_{ik}.
\]

The exclusion restriction demands that $\Bar{Y}_{k,-j}$ and $\varepsilon_{ij}$ be independent of each other. The instrument $\Bar{Y}_{k,-j}$ equals the average value of the dependent variable in some of the other structural equations: those governing whether player $k$ engages in toxic speech in the matches in which they were not paired with player $j$. It follows that the corresponding error terms enter the value of our instrumental variable. Therefore, for our instrumental variable to satisfy the exclusion restriction, we must assume that these error terms are independent of $\varepsilon_{ij}$. A sufficient condition for this is to presume that the error terms for two players in two matches in which they were not paired together are independent:
\begin{multline*}
    \varepsilon_{ij} \perp\!\!\!\perp \varepsilon_{\ell k} \text{ for all matches } i, \ell \text{ in which} \\
    \text{players } j \text{ and } k \text{ are paired.}
\end{multline*}

In interpreting our findings, we must recognize that our identification strategy yields an estimate of the local average treatment effect (LATE) for ``compliers'' \cite{wooldridge_2010}. Therefore, our estimates reflect the effect of teammates' toxic behavior on these specific players. In our context, the compliers are players whose teammates engaged in toxic behavior because they were more likely to behave that way in matches with others and, consequently, exogenously more likely to do so in the current game. This definition excludes players who intentionally use toxic language to provoke reactions from their teammates, for example. If the effect of exposure to toxic language were heterogeneous, the LATE may not accurately reflect the average treatment effect for the entire player population.

\subsection{Generalization to Battle Royale Trios and Quads Game Modes}

So far, our description of the methodology has focused on the BR Duos game mode, in which each player has a single teammate. We now explain how to generalize this approach to game modes in which players have more than one teammate, specifically, the Trios and Quads game modes. In this case, our structural model of players' behavior is defined as follows:
\[
    Y_{ij} = \alpha_{j} + \beta \times \frac{\sum_{k = 1}^{K} Y_{ik}}{K} + \varepsilon_{ij}.
\]

In this equation, $K$ represents the number of teammates player $j$ has. This model is a straightforward extension of the structural model defined above, in which the right-hand-side variable is the share of teammates who engage in toxic speech. In this equation, the coefficient $\beta$ reflects the effect of all of a player's teammates using toxic language on their probability of engaging in similar behavior.

The generalized model suffers from a causal identification problem analogous to the one described above, requiring a similar identification strategy. We instrument the variable representing the share of a player's teammates who use toxic speech with the average probability that each engages in such behavior in other matches in which they are not paired with the player but may still be matched with each other. This instrument satisfies the exclusion restriction. We only consider observations for which we can compute the instrument's value for all a player's teammates since the model would be underspecified if we lacked the value of the instrumental variable for some teammates.

\subsection{Estimation} 

Our regression models include player-specific intercepts, formally called fixed effects, capturing the innate tendency of players to exhibit outcomes of interest. Estimation of these fixed effects is computationally expensive. Therefore, analysts frequently resort to ``down-sampling,'' which consists of sampling a computationally convenient number of observations and estimating the model with fixed effects only for those. This leads to a lower statistical accuracy. However, we do not need to compute them explicitly, especially since these fixed effects are not of primary interest to our analysis. The reason for including them in our model is to absorb time-invariant variables that affect individual players' propensity to display the outcomes of interest. This is particularly important if there is a correlation between a player's inherent tendency to engage in toxic speech and their teammates'. Instead of explicitly estimating fixed effects, we can achieve the same end by demeaning the values of the dependent, independent, and instrumental variables for all players at the individual level \cite{greene_2018}. After doing so, we can estimate the coefficients $\beta$ through the standard 2SLS estimation procedure.

We restrict our analysis to observations for which the following two conditions are met: (i)~we observe the teammate play at least one match with another player so that we can compute the instrument's value, and (ii)~we observe the player participate in at least two matches so that we can demean the values of the dependent, independent, and instrumental variables for that player. These restrictions lead to some minor attrition.

\section{Results}

\begin{table*}
    \centering
    \caption{Estimates of the Effect of Teammates' Toxic Speech on a Player's Probability of Employing Similar Language}
    \label{tab:BR_Duos}
    \makebox[\linewidth][c]{\begin{tabular}{lcccccccc}
        \hline
        \hline
        & & & & & & & \\[-1.4ex]
        & \multicolumn{2}{c}{\multirow{2}{*}{\textbf{Duos Mode}}} & \multicolumn{2}{c}{\textbf{Duos Mode}} & \multicolumn{2}{c}{\multirow{2}{*}{\textbf{Quads Mode}}} & \multicolumn{2}{c}{\multirow{2}{*}{\textbf{All Modes}}} \\
        & & & \multicolumn{2}{c}{\textbf{Algorithmic Pairs}} & & & & \\
        & & & & & & & \\[-1.4ex]
        \cline{2-9}
        & & & & & & & \\[-1.4ex]
        & OLS & 2SLS & OLS & 2SLS & OLS & 2SLS & OLS & 2SLS \\
        & & & & & & & \\[-1.4ex]
        \hline
        & & & & & & & \\[-1.4ex]
        Estimate & 0.0375 & 0.0737 & 0.0185 & 0.0291 & 0.0596 & 0.0650 & 0.0420 & 0.0757 \\
        Standard Error & (0.001) & (0.0211) & (0.003) & (0.0118) & (0.002) & (0.0091) & (0.001) & (0.0088) \\
        & & & & & & & \\[-1.4ex]
        \hline
        & & & & & & & \\[-1.4ex]
        Multiple of Average Probability of Toxic Speech & 14.2 & 29.6 & 18.2 & 28.6 & 23.9 & 26.1 & 16.9 & 30.3 \\[1.2ex]
        \hline
        \hline
    \end{tabular}}
\end{table*}

Table \ref{tab:BR_Duos} presents estimation results for all four sets of observations described above. The table presents the estimated values of the coefficients reflecting the causal effect of all a player's teammates engaging in toxic speech on that player's probability of engaging in such behavior, along with their heteroskedasticity-robust standard errors. To provide readers with a sense of this effect's relative magnitude, we also express the value of the coefficients as a multiple of the average probability that players use toxic speech. Finally, for comparison, we present OLS estimates, which lack a causal interpretation, along with those of our 2SLS estimation strategy, to which we ascribe a causal interpretation.

In the BR Duos mode, our analysis reveals that when a player's teammate engages in toxic speech, it causes a 7.37 pp. increase in the probability that the player also exhibits such behavior. This effect equals 29.6 times the average likelihood of a player engaging in toxic speech. This means that, all else equal, a player whose teammate engages in toxic speech is 29.6 more likely to engage in such behavior than the average player. This effect is statistically significant at the 99\% confidence level. OLS estimates suggest that the probability of a player engaging in toxic speech is 3.75 pp. higher when their teammate does, a magnitude roughly half that of the estimated causal effect.

For the algorithmically matched pairs in the BR Duos game mode, our analysis indicates that when their teammate engages in toxic speech, it causes a 2.91 pp. increase in the probability that a player exhibits such behavior. This effect equals 28.6 times the average likelihood of a player engaging in toxic speech. This effect is statistically significant at the 95\% confidence level. In comparison, OLS estimates suggest that the probability of a player engaging in toxic speech is 1.85 pp. higher when their teammate does, a magnitude roughly half that of the estimated causal effect.

In the BR Quads mode, our analysis indicates that all their teammates indulging in toxic speech causes a 6.5 pp. increase in the probability that a player engages in such behavior.\footnote{The causal effect of a single teammate using toxic language can be computed by multiplying the coefficient by the portion of teammates it represents, that is, by one-third.} This effect equals 26.1 times the average likelihood of a player engaging in toxic speech. This effect is statistically significant at the 99\% confidence level. In comparison, OLS estimates suggest that the probability of a player engaging in toxic speech is 5.96 pp. higher when all their teammate does, which is close but still lower than the estimated causal effect.

For all BR game modes pooled together, our findings indicate that when all their teammates engage in toxic speech, it causally increases by 7.57 pp. the probability that a player participates in such behavior. This effect equals 30.3 times the average likelihood of a player engaging in toxic speech. This effect is statistically significant at the 99\% confidence level. OLS estimates suggest that the probability of a player engaging in toxic speech is 4.20 pp. higher when all their teammate does, which is slightly more than half the magnitude of the causal effect.

\section{Discussion and Conclusion}

Our analysis reveals that all else equal when a player's teammates engage in toxic speech, it increases the player's probability of exhibiting similar behavior by 26.1 to 30.3 times the average player's likelihood of engaging in toxic speech. This implies that toxic behavior spreads among teammates, amplifying the effects of a single player's misconduct. These findings underscore the viral nature of toxicity, particularly toxic speech, in competitive online video games.

The overall effect exerted by a player's teammates on their probability of engaging in toxic speech is consistent across the BR Duos and Quads game modes. This implies that the causal effect of a single teammate engaging in toxic speech is lower in the Quads game mode compared to the Duos mode. In other words, the presence of more teammates diminishes the influence of a single teammate engaging in toxic speech. This is plausible since a player has an inherently lower probability of interacting with any specific teammate when they have many, as in the Quads game mode, compared to when they only have one, as in the Duos game mode.

Our findings also suggest that the endogeneity induced by the reflection problem is less pronounced in the Quads game mode, as evidenced by the smaller gap between OLS and 2SLS estimates compared to the Duos mode. This is also sensible, considering that the effect of a single teammate participating in toxic speech is lower in the Quads game mode, and the likelihood of any two teammates interacting decreases as team size grows. As a result, the ``reflection'' of one teammate's effect on a player's likelihood of engaging in toxic speech is lower, resulting in less endogeneity.

While we cited conformity as the primary justification for peer effects, they may also be driven by homophily. Homophily is characterized by players' tendency to assemble teams with others sharing a similar inclination for engaging in toxic speech or with whom they intend to engage in such behavior \cite{CHARROIN2022618}. Similar to how the reflection problem makes it more likely that players engage concurrently in toxic behavior, homophily increases the probability of players concurrently engaging in toxicity.

Our model includes fixed effects to control for the innate inclination of players and their teammates towards toxic speech. Therefore, our estimates reflect how much more likely than average a player is to use toxic language when paired with teammates who are exogenously more prone to such behavior than their average teammate. Additionally, our instrumental variable estimation strategy is designed to neutralize the effect of players intentionally forming teams to engage in toxic behavior. Nonetheless, to address potential apprehensions about our estimates being inadvertently influenced by the endogenous formation of teams, we can leverage the distinction between players who joined a match in endogenously formed ``parties'' and those who joined alone and were algorithmically matched with a teammate. Specifically, we estimated peer effects using the subset of algorithmically matched pairs in the Duos game mode. While the absolute magnitude of peer effects in algorithmically matched pairs is lower compared to that of all Duos, these effects' magnitude relative to the average probability that a player in an algorithmically matched pair engages in toxic speech is the same as that observed across all observations in the Duos game mode. This implies that the magnitude of estimated peer effects is unaltered by homophily.

Our work expands on and contributes to the literature on toxicity in competitive online video games by uncovering its viral nature. Rather than relying on survey data, as many studies have done, we take a distinct yet complementary approach, focusing on observational causal inference with gameplay data. This approach mitigates the various biases associated with players' self-reported behavior and perceptions, though it introduces greater analytical complexity. We formulate an estimation strategy to measure the causal effects of exposure to toxicity. This approach could be applied to further document the effects of toxicity on player performance \cite{turkay_2020}, retention \cite{kowert_kilmer}, and other key outcomes, thereby strengthening previous research in this area.

Admittedly, our analysis does not account for whether players were exposed to toxic speech. Instead, we consider how teammates engaging in toxic speech affects a player's behavior, irrespective of whether that player was exposed to that speech. However, if a player's audio chat is disabled, they cannot hear their teammates' toxic speech. In this case, it cannot plausibly affect their behavior. On the whole, this results in conservative estimates because the effect of direct exposure to toxic speech is necessarily greater than the effect of teammates simply engaging in toxic speech, as the latter is a prerequisite for the former. In econometric terms, our estimates can be interpreted as reflecting the effect of the ``intent to treat,'' where the treatment is exposure to toxicity, rather than the effect of the treatment itself.

In conclusion, we suggest that future research investigate in priority how the spread of toxic speech to a player's teammates varies across different contexts. Understanding these differences would be invaluable, as it would allow resources to be allocated to the contexts where the virality of toxicity and, by extension, its adverse effects are the most pronounced. Exploring forms of toxicity beyond language and speech also offers a compelling direction for future research.

\section*{Acknowledgment}

The authors thank Andrea Boonyarungsrit, Grant Cahill, MJ Kim, Rafal Kocielnik, Jonathan Lane, Zhuofang Li, Gary Quan, Deshawn Sambrano, Feri Soltani, Carly Taylor, and Michael Vance for their assistance and support in preparing this article. The opinions expressed by the authors do not represent the views of Activision\textregistered.

\ifCLASSOPTIONcaptionsoff
  \newpage
\fi



\bibliographystyle{IEEEtran}
\bibliography{bibliography}
%

\end{document}